\documentclass[oneside,a4paper,12pt,shownumbers,manualsort]{article}
\usepackage{latexsym}
\usepackage{euscript}
\usepackage{epsfig,amsmath,amssymb}
\usepackage{slashed}
 
\renewcommand{\hat}[1]{{#1}}
\renewcommand\({\left(}
\renewcommand\){\right)}
\renewcommand\[{\left[}
\renewcommand\]{\right]}

\newcommand{\dd}{{\rm d}}
\newcommand{\e}{{\rm e}}
\def\be{\begin{equation}}
\def\ee{\end{equation}}
\def\bea{\begin{eqnarray}}
\def\eea{\end{eqnarray}}

\newcommand\GeV{\,\mbox{GeV}}

\newcommand\mpl{M_{\rm pl}}

\newcommand\mcL{\mathcal L}
\newcommand\mcA{\mathcal A}
\newcommand\mcM{\mathcal M}

\long\def\symbolfootnote[#1]#2{\begingroup%
\def\thefootnote{\fnsymbol{footnote}}\footnote[#1]{#2}\endgroup} 

 \large\normalsize

\begin{document}

\begin{center}

{\Large \bf Majorana zero modes}

\vspace*{7mm}
{\ Rachel Jeannerot$^{a}$\symbolfootnote[1]{{E-mail:jeannerot@lorentz.leidenuniv.nl}} and Marieke Postma$^{b}$}\symbolfootnote[2]{E-mail:mpostma@nikhef.nl}
\vspace*{.25cm}

${}^{a)}${\it Instituut-Lorentz for Theoretical Physics,
Niels Bohrweg 2, 2333 CA Leiden, The Netherlands}\\
\vspace*{.1cm} 
${}^{b)}${\it NIKHEF, Kruislaan 409, 1098 SJ Amsterdam,
The Netherlands}


\end{center}

\begin{abstract}
  {We derive the zero mode solutions for a Majorana fermion in the
    background of a cosmic string and contrast it with the zero mode
    solution for a (neutral) Dirac fermion. A Majorana zero mode
    carries no vector or axial charge, and it cannot be bosonised.  We
    study the implications for vorton formation and stability.  In the
    massless limit stability of the zero mode is guaranteed by
    energy-momentum conservation.  However, zero modes obtain an
    effective mass on string loops. It is found that the conditions
    under which current formation can be effective are exactly those
    for which zero mode decay is most likely to occur.}
\end{abstract}


\section{Introduction}

The existence of Majorana mass terms is particularly important for
particle-physics phenomenology in the context of neutrino physics
\cite{majorana}.  Indeed, the see-saw mechanism \cite{see-saw}, which
can naturally explain the smallness of the observed left-handed
neutrino masses \cite{observed}, relies on the existence of a Majorana
mass term for a standard model singlet, the right-handed
neutrino. Right-handed neutrinos and the see-saw mechanism are a
prediction of unified theories which are left-right symmetric, such as
Patti-Salam and SO(10) grand unified theories (GUTs).  After breaking
of the gauged $U(1)_{B-L}$ symmetry a Majorana mass for the the
right-handed neutrino is generated through the Higgs mechanism.

In the context of cosmology, GUTs predict the existence of cosmic
strings \cite{Kibble,laz}. If the Higgs field which form the string breaks
$U(1)_{B-L}$, $B-L$ cosmic strings are produced~\cite{lept,PRD}. They can be
important for baryogenesis \cite{lept,sahu}. Cosmic strings can also
contribute to the cosmic microwave background anisotropies and they
produce gravitational waves \cite{ShelVil}. Further, fermions coupling
to the string forming Higgs field can be trapped on the string,
travelling along the string at the speed of light~\cite{jackiw}.  Such
fermionic zero modes can give rise to superconducting
strings~\cite{Witten}.  Conserved fermionic currents can stabilise
cosmic string loops against gravitational collapse, resulting in the
formation of vortons which are potentially cosmologically
catastrophic~\cite{vortons}. It was shown, however, that if there are
fermionic zero modes which travel in opposite directions on the string
they could scatter with each other, thereby strongly reducing the
current \cite{Barr,RLDavis}. Hence, string loops with chiral currents
are the best vorton candidates. In an anomaly free theory, chiral zero
modes have to be neutral; they can either be Dirac or Majorana.

In this paper we will be interested in chiral zero modes and in
particular in zero modes for Majorana fermions.  Such zero modes arise
naturally in $B-L$ cosmic strings, as the $U(1)_{B-L}$ breaking Higgs
field gives a Majorana mass to the right-handed neutrino \cite{lept}.
They can produce the observed lepton asymmetry of the universe,
independently of the reheating temperature after
inflation~\cite{lept,PRD}. We will discuss the difference between
Majorana and Dirac zero modes. Neutral Dirac currents are persistent
by virtue of a conserved quantum number, whereas Majorana currents are
persistent for kinematical reasons.  The difference between the two
currents shows up when the zero-mode obtains an effective mass. Decay
and annihilation of Dirac zero modes can be suppressed or forbidden by
virtue of lepton number conservation; no such suppression exists for
Majorana fermions. We apply our results to vortons. Neutrinos zero
modes were discussed before in a slightly different context
\cite{vachaspati}.

This paper is organised as follows.  In the next section we will give
a brief introduction to Majorana and Dirac spinors.  In
section~\ref{s:zeromode} we present the zero mode solutions for both
the Dirac and Majorana spinors in the presence of a Nielsen-Olesen
string.  We quantise the string in section~\ref{s:quantization}, and
compute the conserved currents.  Moreover, a description of the
effective theory in 2D is given.  We apply our results to vortons in
section~\ref{s:vortons}. Both current condensation and stability is
discussed, highlighting the difference between Dirac and Majorana
currents. We end with some concluding remarks.

\section{Dirac and Majorana spinors}
\label{s:dirac_maj}

In the early Universe when particles are still massless, there is no
distinction between Dirac and Majorana fermions. All fermions and
anti-fermions can be described by two-component Weyl spinors. In grand
unified theories fermions and anti-fermions are in the same
representation and therefore must be described by spinors with the
same chirality. By convention, we use left-handed Weyl
spinors.\footnote{Our notation and conventions are detailed in the
appendix.} Weyl spinors in 4-dimensions have two on-shell degrees of
freedom.

Fermions get masses via the Higgs mechanism. The mass term involves a
coupling between a fermion bilinear and a scalar field $\phi$ which
acquires a non-vanishing vacuum expectation value (VEV). The mass term
must be invariant under all symmetries of nature, and in particular,
it must be invariant under Lorentz transformations.

Given two left-handed Weyl spinors $\psi_L$ and $\chi_L$, the Lorentz
invariant bilinear which can be constructed is of the form
\be
\chi_L^T \tilde{C} \psi_L 
+ \overline{\chi_L} \tilde{C}  \overline{\psi_L}^T .
\label{eq:mass}
\ee
If the Weyl spinors transform under some symmetry, the above term is in
addition gauge invariant if $\psi_L$ and $\chi_L$ are charge conjugate
of each other, i.e., if $\chi_L =\psi_L^c$. The bilinear term given in
Eq.~(\ref{eq:mass}) then becomes:
\be
(\psi^c_L)^T \tilde{C}  \psi_L + h.c. = (\tilde{C} 
\overline{\psi_R}^T)^T \tilde{C} \psi_L + h.c. 
= \overline{\chi_R} \tilde{C}^T \tilde{C} \chi_L + h.c.
= \overline{\psi_R}  \psi_L + \overline{\psi_L} \psi_R.
\label{eq:dir}
\ee
This leads to the standard Dirac mass term
\be
{\mathcal L}_D = 
- \overline{\psi_R}  m_D \psi_L -  \overline{\psi_L}  m_D^\dagger \psi_R
\ee
with $m_D = \lambda \phi$ and $\lambda$ a Yukawa coupling. The two
Weyl spinors $\psi_L$ and $\psi_R = \tilde{C}\overline{\psi^c_L}^T$
combine to make a Dirac spinor 
\be
\psi_D = \psi_L + \psi_R,
\ee
and if $m_D$ is real, the mass term can be written as ${\mathcal
L}_{D} = - m_D \overline{\psi_D} \psi_D$.  A Dirac spinor has on shell
four real degrees of freedom corresponding to particle/antiparticle
and positive/negative helicities.  The Dirac mass term is invariant
under a global $U(1)$ transformation; this implies the existence of a
quantum number (such as lepton number) which enables to distinguish
particles from antiparticles. A Dirac mass term can be written for
both neutral and charged particles.

Now starting with a single neutral Weyl spinor, i.e., a particle which
does not carry any quantum number under any unbroken symmetry, we can
also construct, by setting $\chi_L = \psi_L$ in Eq.~(\ref{eq:mass}),
the following Lorentz invariant:
\be 
\psi_L^T \tilde{C} \psi_L  + \overline{\psi_L} \tilde{C}  \overline{\psi_L}^T 
= \overline{\psi_R^c}  \psi_L + \overline{\psi_L} \psi_R^c.
\label{eq:maj}
\ee
We can then write down what is called a Majorana mass term 
\be
{\mathcal L}_{M} = - {1\over 2}  \overline{\psi_R^c}  m_M  \psi_L 
- {1\over 2} \overline{\psi_L}  m_M^\dagger \psi_R^c,
\label{eq:m_maj}
\ee
where $m_M = \lambda \phi$ and the factor half is conventional.
Analogously to the Dirac spinor, a 4-component Majorana spinor can
then be defined 
\be 
\psi_M = \psi_L + \tilde{C} \overline{\psi_L^c}^T 
= \psi_L + \psi_R^c
= \psi_{M,L} + \psi_{M,R}.
\label{eq:psi_M} 
\ee 
If $m_M$ is real, the mass term simplifies to ${\mathcal L}_{M} = -
{1\over 2} m_M \overline{\psi_M} \psi_M$. A Majorana mass term is not
invariant under a global $U(1)$ transformation, and there is no
quantum number which allows to distinguish particles from
antiparticles (for instance, lepton number cannot be defined for a
Majorana neutrino).  A Majorana spinor is made of a single Weyl spinor
and it has on shell two real degrees of freedom corresponding to the
two helicity states. It satisfies the Majorana condition
\be 
\psi_M^c = \psi_M,
\label{eq:maj_cond}
\ee
that is, a Majorana particle is its own antiparticle.

Majorana masses are particularly important in understanding the
smallness of the neutrino masses via the see-saw mechanism
\cite{see-saw}. Indeed, the latter relies on the existence of a
superheavy Majorana mass term for the right-handed neutrino on top of
a standard Dirac mass term which arises from the coupling of both left
and right-handed neutrinos to the standard model Higgs.  In principle
there can also be a small Majorana mass for the left-handed
neutrino. There are then two distinct mass eigenstates which are both
Majorana: a state with a tiny mass which is a superposition of mostly
left-handed neutrino with a small admixture of the right-handed
neutrino, and a superheavy state which is mostly composed of the
right-handed neutrino.  In the context of topological defects, we will
be interested in a Majorana mass for the right-handed neutrino which
arises as a consequence of gauge symmetry breaking.

\begin{table}[t]
\begin{center}
\begin{tabular}{| c | c | c | c |}\hline
Dirac & $\phi$ & $\psi_L$ & $\chi_L$ \\\hline\hline
X & $r$ & $n$ &  $p=-n-r$ \\\hline
Q & 0 & $q$ & $-q$   \\\hline
\end{tabular} \hspace{1cm}
\begin{tabular}{| c | c | c | }\hline
Majorana & $\phi$ & $\psi_L$ \\\hline\hline
X & $r$ & $-r/2$  \\\hline
Q & 0 & 0  \\\hline
\end{tabular}
\end{center}
\caption{Charges of the Higgs field $\phi$ and of the fermion fields
$\psi_L$ and $\chi_L$, which acquire respectively Dirac and Majorana
mass by coupling with $\phi$, under the broken $U(1)_X$ and any
unbroken $U(1)_Q$. Here $r,n\neq 0$, but $q$ can either be zero or
non-zero.}
\end{table}

In summary, massive charged fermions are always Dirac fermions whereas
neutral fermions can either be Dirac or Majorana.  Which mass terms
can be constructed depends on the quantum numbers of the fermions and
the scalar fields of the model.  To be explicit, consider a $U(1)_X
\times U(1)_Q$ gauge symmetry, possibly embedded in a larger gauge
group.  The $U(1)_X$ symmetry is broken by the non-vanishing vacuum
expectation value of a complex Higgs field $\phi$, while $U(1)_Q$
remains unbroken. Given two left-handed Weyl fermions $\psi_L$ and
$\chi_L$, Dirac and Majorana mass terms involving the bilinear term
given in Eq.~(\ref{eq:mass}) multiplied by the Higgs field $\phi$ can
be constructed if the charges of the fields are as given in
Table~1. Let's call $X$ and $Q$ the charges of the fields under
$U(1)_X$ and $U(1)_Q$ respectively.  $\phi$ breaks $U(1)_X$ and keeps
$U(1)_Q$ unbroken, hence $X(\phi) = r \neq 0$ and $Q(\phi)=0$. Let
$X(\psi_L) = n$ and $X(\chi_L) = p$. Then the Yukawa term is gauge
invariant only if $n + p +r = 0$ and $Q(\psi_L) = - Q(\chi_L)$. We
distinguish three cases:

1. If $Q(\psi_L) = - Q(\chi_L) \neq 0$, we have a Dirac mass term. The
spinors $\psi_L$ and $\chi_L$ combine to form a Dirac spinor which has
four on shell degrees of freedom.  It describes a charged fermion.

2. If $Q(\psi_L) = 0$ and $n = -r/2$, we have a Majorana mass. A
Majorana spinor is formed out of $\psi_L$, it has two on-shell degrees
of freedom. It describes a Majorana fermion, a particle which is its
own antiparticle.

3. Now if $Q(\psi_L) = 0$ and $n \neq r/2$, we have a Dirac mass term
for a neutral particle. In this case, $\psi_L$ and $\chi_L$ combine to
form a Dirac spinor. Note that in this case always a global charge can
be defined which distinguishes particles from antiparticles.  If there
is a second Higgs field with charge $X = -2 X(\psi)$, respectively $X
= -2 X(\chi)$, a Majorana mass term can be constructed for $\psi$
(respectively $\chi$). There are then two distinct mass eigenstates
which are Majorana.

\section{Chiral currents} 
\label{s:zeromode}

\subsection{Zero modes solution}

In this section, we write down the fermionic equations of motion in
the background of a cosmic string.  We give the zero mode
solutions for both Dirac and Majorana fermions, and compare the
results.  We only discuss Abelian cosmic strings; our results can
be generalised straightforwardly to non-Abelian strings.

Consider a $U(1)_X$ gauge symmetry which is broken down to the
identity by a Higgs field $\phi$, when the latter acquires a
non-vanishing VEV. Since $\pi_1(U(1)_X) \neq 0$, cosmic strings form
at the $U(1)_X$ breaking scale according to the Kibble
mechanism~\cite{Kibble}. For a straight infinite cosmic string lying
along the $z$-axis, the Higgs field $\phi$ and gauge field $A$ have
the form
\bea
\phi & = & \phi_0 f(r) e^{i n \theta},  \\
A_{\theta} & = & - n \frac{g(r)}{ e r }, \hspace{1cm}  A_{z} = A_{r} = 0 ,
\eea
with $\phi_0$ the Higgs VEV in the vacuum, and $n \in N$ the string's
winding number. In the following we will concentrate on the minimum
energy configuration corresponding to $|n|=1$.  The profile functions
$f(r)$ and $g(r)$ must satisfy the following boundary conditions
\bea && f(0) = g(0) = 0, \\
&& \mathop{\lim}\limits_{r \, \rightarrow \infty} f(r) = 
\mathop{\lim}\limits_{r \, \rightarrow \infty} g(r) = 1. \eea
The exact form of the functions $f(r)$ and $g(r)$ depends on the Higgs
potential~\cite{ShelVil}, which we do not specify.\\

We first discuss the well known zero mode solution \cite{jackiw} for a
Dirac fermion getting its mass by coupling to $\phi^*$~\cite{Witten}.
The Lagrangian is 
%
\be
{\mathcal L}_D = \overline{\psi_L} i \gamma^\mu D_\mu \psi_L
+ \overline{\psi_R} i \gamma^\mu D_\mu \psi_R
- \lambda \phi^* \overline{\psi_R} \psi_L 
- \lambda \phi \overline{\psi_L} \psi_R,
\label{lag_D}
\ee
with $D_\mu = \partial_\mu+ i q A_\mu$, $\lambda$ the Yukawa coupling,
and $q$ the charges of $\psi_{L,R}$ under $U(1)_X$.  Note that $q_\phi
-q_L+q_R =0$; in general $q_L$ and $-q_R$ can be different.  The
equations of motions are:
\bea
i\gamma^\mu D_\mu \psi_L &=& \lambda \phi \psi_R, \nonumber \\
i\gamma^\mu D_\mu \psi_R &=& \lambda \phi^* \psi_L. 
\eea
From Jackiw and Rossi \cite{jackiw}, and an index theorem
\cite{Weinberg}, we know that there is one normalisable zero mode
solution $\psi^0$ (a static solution) which is $z$ and $t$
independent. The zero mode solution are eigenvectors of the projection
operator~\cite{Witten}
\be
\gamma^0 \gamma^3 \psi^0 = n \psi^0
\label{proj}
\ee
with $n=1$ for a vortex and $n=-1$ for an anti-vortex. With our choice
of gamma matrices (see the appendix) the zero mode solutions are
\be
\psi^0_{n=+1} =  
\(\begin{matrix}
0 \\
\alpha_+ (r,\theta)  \\
\beta_+ (r,\theta) \\
0
\end{matrix}\), 
\qquad \qquad
\psi^0_{n=-1} = 
\(\begin{matrix}
  \alpha_{-} (r,\theta)  \\
0 \\
0 \\
\beta_{-}(r,\theta) 
\end{matrix}\),
\label{psi_0}
\ee
for a vortex and an anti-vortex respectively.  The wave functions
$\alpha_\pm(r,\theta)$ and $\beta_\pm(r,\theta)$ are of the form $f(r)
e^{il\theta}$ with l an integer and $f(r)$ a real function localised
on the string.  We define
\bea 
\alpha_\pm &=& \int r \dd r \dd \theta \, \alpha_\pm(r,\theta), \nonumber \\
\beta_\pm &=& \int r \dd r \dd \theta \, \beta_\pm(r,\theta).  
\label{ab}
\eea 
The ratio $\alpha_\pm/\beta_\pm$ is determined by the equations of
motion, more specifically, by the charges $q_L$ and $q_R$.  Further
the sum $(\alpha_\pm^2 + \beta_\pm^2)$ is set by the spinor
normalisation.  Hence there is no freedom in choosing the (relative)
magnitudes of $\alpha_\pm$ and $\beta_\pm$.  We normalise
\be
{\psi^0_n}^\dagger \psi^0_n = 1 
\qquad \Longrightarrow \qquad
\alpha_\pm^2 + \beta_\pm^2 = 1
\label{normalization}
\ee

The 4-dimensional solutions of the Dirac equation are of the
form\footnote{For a Dirac particle, let's call it $\eta$, getting its
mass by coupling to $\phi$ instead of $\phi^*$, the zero mode
solutions are $\eta^0_{n=1} = \psi^0_{n=-1}$ and $\eta^0_{n=-1} =
\psi^0_{n=1}$.  It follows that for a vortex a fermion zero mode due
to Yukawa coupling to $\phi$ moves in opposite direction as a fermion
zero mode due to Yukawa coupling to $\phi^*$.}
\be
\psi_D = \psi^0_n \e^{\pm i E(t-nz)},
\label{zm_D}
\ee
where we have set $E = k = k_z$, with $k > 0$, for a string aligned
along the $z$-axis. The zero-modes are massless excitations which
travel at the speed of light along the string. Fermions trapped on a
vortex (anti-vortex) travel in the $+z$ $(-z)$ direction.  Note that
both particles and anti-particles move in the same direction along the
string, the direction being determined by whether the Yukawa coupling
in Eq.~(\ref{lag_D}) is to a vortex or anti-vortex.  The projection
operator in Eq.~(\ref{proj}) singles out one helicity state
(determined in terms of $\alpha_+$ and $\beta_+$ or $\alpha_-$ and
$\beta_-$). The zero mode solution has thus only two real degrees of
freedom which correspond to particle and anti-particle.  It is
effectively a 2-dimensional Weyl fermion.

Now consider a Majorana fermion getting its mass by coupling to
$\phi^*$.  The Lagrangian reads
\be
{\mathcal L}_M =  
\overline{\psi_L}  i \gamma^\mu D_\mu \psi_L 
- \frac12 \lambda \phi \overline{\psi_R^c} \psi_L
- \frac12 \lambda \phi^* \overline{\psi_L} \psi_R^c 
\label{lag_M}
\ee
The equations of motions are
\bea
i\gamma^\mu D_\mu \psi_L &=&  \lambda \phi \psi_R^c, \nonumber \\
i\gamma^\mu D_\mu \psi_R^c &=&   \lambda \phi^* \psi_L. 
\label{eom_maj}
\eea
Normalising the $U(1)_X$ charge of $\phi$ to unity, the charges of
$\psi_L$ and $\psi_R^c$ are $q_L = -q_R^c=-1/2$.  The two equations of
motion are not independent, rather the second equation in
Eq.~(\ref{eom_maj}) can be obtained from the first one by charge
conjugation. The solution to the equation of motion is 
\be
\psi_M^{\rm zm} = {1\over \sqrt{2}} (\psi^0_n(r,\theta) \e^{-iE(t-nz)} 
+  \psi^{0*}_n(r,\theta) \e^{+iE(t-nz)})
\ee
with $\psi^0_n$ as given in Eq.~(\ref{psi_0}) with the additional
constraint that
\be
\alpha_\pm(r,\theta) = \mp \beta_\pm(r,\theta).
\label{eq}
\ee
The Majorana solution is just a linear combination of the Dirac
solutions given in Eq.~(\ref{zm_D}) which satisfies the reality
constraint of Eq.~(\ref{eq:maj_cond}). There is only one solution
corresponding to one real degree of freedom, which is half the number
of degrees of freedom of the Dirac solution. The zero mode is
effectively a 2 dimensional Majorana-Weyl fermion A consequence of
Eq.~(\ref{eq}) is that both the vector and axial currents are zero.
Nevertheless there is energy-momentum flowing along the string.

\subsection{Quantisation and conserved currents}
\label{s:quantization}

In this section we quantise the Dirac and Majorana zero mode fields,
and evaluate the conserved charges and energy-momentum.

As before, we start with the Dirac spinor. We use the classical
wave solutions to the Dirac equation, Eqs.~(\ref{psi_0},~\ref{zm_D}),
to write the quantised field 
\be
\hat{\psi}_n = \int_{0}^\infty \frac{\dd k}{(2\pi)} 
\( 
\hat{a}(nk) \psi^0_n e^{-i k(t-n z)} + 
\hat{b}^\dagger(nk) \psi^0_n e^{i k(t-n z)} 
\),
\ee
with as before $k \equiv k_z$. Note that $k>0$ always. Thus
$\hat{a}(k<0)$ are annihilation operators for states moving in the
$-z$ direction, which can only be exited for an anti-vortex ($n=-1$).
Likewise $\hat{a}(k>0)$ are annihilation operators for states moving
in the $+z$ direction, which can only be exited for a vortex
($n=1$). The wave function $\psi^0$ is $k$-independent by virtue of
separation of variables in the equation of motion; the normalisation
is also momentum independent.

Imposing the equal time anti-commutation relations
\be
\{\hat{\psi}(x,y,z),\hat{\psi}^\dagger(x,y,z')\} 
= \delta(z-z') \psi_0 \psi_0^\dagger,
\ee
it follows that the operators $\hat{a},\, \hat{b}$ satisfy the
anti-commutation relations $\{\hat{a}(k),\hat{a}^\dagger(k')\}
=\{\hat{b}(k),\hat{b}^\dagger(k')\} = 2 \pi \delta(k-k')$.  One can
construct a Fock space in the usual way~\cite{Ringeval1}.

The vector current $j^\mu = \bar{\psi} \gamma^\mu \psi$ and axial
current $j^{\mu5} = \bar{\psi} \gamma^\mu \gamma^5 \psi$ are
identically zero for $\mu=1,\,2$.  The non-zero components for $\mu =
0,3$ are related, $j^3 = n j^0$ and $j^{35} = n j^{05}$, by virtue of
Eq.~(\ref{proj}).  The conserved charges corresponding to the vector
and axial currents, $Q = \int \dd^3 x \langle \, :\!j^0\!: \, \rangle $
and $Q^5 = \int \dd^3 x \langle \, :\! j^{05}\!: \, \rangle$
respectively, are
\bea
Q &=& \int_{0}^\infty \frac{\dd k}{2\pi} 
\langle \hat{a}^\dagger(k) \hat{a}(k) 
-  \hat{b}^\dagger(k) \hat{b}(k) \rangle
\nonumber \\ 
&=& N - \bar{N},\\
Q^5 &=& \int_{0}^\infty \frac{\dd k}{2\pi}
\langle
(-\alpha^2 + \beta^2) \hat{a}^\dagger(k) \hat{a}(k)
+ (\alpha^2 - \beta^2) \hat{b}^\dagger(k) \hat{b}(k)
\rangle
\nonumber \\ 
&=& (\alpha^2-\beta^2) (-N + \bar{N}).
\label{Q}
\eea
where, to avoid notational cluttering, we have dropped the subscript
$n$ to indicate whether it concerns a vortex or anti-vortex.  Here
$:\;:$ denotes normal ordering, i.e., putting all annihilation
operators to the left; this procedure automatically subtracts the zero
point contributions.

The energy momentum tensor 
\be
T_{\mu \nu} = \bar{\psi} i \gamma_\mu \partial_\nu \psi 
- \eta_{\mu \nu} \bar{\psi} i \gamma^\sigma \partial_\sigma \psi
\label{T}
\ee
is likewise only non-zero for $\mu,\nu = 0,3$.  The various components
are related: $T_{00} = T_{33}= -n T_{03}$ and lead to the zero mode
contribution to the energy per unit length of the string (respectively
to the tension):
\be
E_F = - T_F  = \int_{0}^\infty \frac{\dd k}{2\pi} 
k \langle  \hat{a}^\dagger(k) \hat{a}(k) 
+  \hat{b}^\dagger(k) \hat{b}(k) \rangle,
\label{H}
\ee

The quantisation of the Majorana field goes analogous.  We write
\be
\hat{\psi}_n = \int_{0}^\infty \frac{\dd k}{(2\pi)} 
\( 
\hat{a}(nk) \psi^0_n e^{-i k(t-n z)} 
+ \hat{a}^\dagger(nk) \psi^0_n e^{i k(t-n z)} 
\),
\ee
supplemented by the anti-commutation relation $\{\hat{a}(k),
\hat{a}^\dagger(k')\} = 2 \pi \delta(k-k')$. The creation and
annihilation operators for the particle and the antiparticle are now
identical. Formally the Majorana zero mode can be obtained from the
Dirac zero mode by setting $\beta_\pm \to \mp \alpha_\pm$ and $\hat{b}(k)
\to \hat{a}(k)$.

The calculation of the conserved currents gives the following.  As for
any Majorana particle, the vector current vanishes (a Majorana mass
term is not invariant under a vector $U(1)$). The axial current is
also zero, by virtue of Eq.~(\ref{eq}): the amount of left-chiral and
right-chiral fields exited for each zero mode is equal.  We thus find
\be
Q = 0, \qquad Q^5 =0
\ee
This can also be found from Eq.~(\ref{Q}), by substituting $\hat{b}(k)
\to \hat{a}(k)$ and $\beta \to \alpha$, as appropriate for a Majorana
spinor.  However, the energy-momentum tensor is non-zero for $\mu,\nu
= 0,3$, as can be seen from Eqs.~(\ref{T},~\ref{H}). This gives a
non-zero contribution to the string energy per-unit-length and the
string tension proportional to the number of Majorana particles
trapped on the string:
\be 
E_F = - T_F = \int_{k>0} \frac{\dd k}{2\pi}
k \langle  \hat{a}(k)^\dagger \hat{a}(k) \rangle.
\ee

Before we go to a discussion of the effective 2 dimensional theory, a
word on notation/nomenclature. In the following we will refer to
$Q$ as particle number, the number of particles minus anti-particles
(it is zero in the Majorana case).  The term excitation number will be
reserved for the total number of excitations, particles plus
anti-particles.  The excitation number does not correspond to a
conserved current.  It can be inferred from energy-momentum.

\subsection{Dimensional reduction}

The zero-mode solutions effectively live in 2 dimensions.  In this
section we discuss how to dimensionally reduce from 4 to 2 dimensions.

Define the matrices
\be
\Gamma^{0\pm} = \frac12 (\pm \gamma^0 + \gamma^3),
\qquad
\Gamma^{1\pm} = \frac12 (\gamma^1 \pm i \gamma^2).
\ee
They satisfy the algebra $\{\Gamma^{i+}, \Gamma^{j-} \} = \delta^{ij}$
and $\{\Gamma^{i\pm}, \Gamma^{j\pm} \} = 0$, and thus act as creation
and annihilation operators. Define the ground state $\zeta$ as the
state which is annihilated by $\Gamma^{i-} \zeta =0$ for all $i$.  We
can then construct all states in the spinor representation by acting with
creation operators on the vacuum.  In 4D these states can be labelled
by $(s_0,s_1)$, with $s_i = \pm 1/2$, corresponding to the state
\be
\zeta^{(s)} =(\Gamma^{0+})^{s_0+1/2} (\Gamma^{1+})^{s_1+1/2} \zeta
\ee
The zero mode solutions $\psi_\pm^0$, see Eq.~(\ref{psi_0}), in this
language are
\bea
\psi^0_+ &=& \alpha_+ (-1/2,-1/2) + \beta_+ (-1/2,+1/2)
\nonumber \\
\psi^0_{-} &=& \alpha_{-} (+1/2,+1/2) + \beta_{-}(+1/2,-1/2).
\eea

Under dimensional reduction from 4 to 2 dimensions
\be
SO(3,1) \to SO(1,1) \times SO(2) \sim SO(1,1) \times U(1)
\ee
the spinor decomposes as $4 \to 2 + 2$.  The $2$'s are spinor
representation of $SO(1,1)$, which are charged under the action of the
global $U(1)$.  We can construct the 2D spinor representation
analogous to the 4D construction.  There is only one pair of
creation/annihilation operators, given by
\be
\tilde{\Gamma}^{0\pm} = \frac12 (\pm \tilde{\gamma}^0 + \tilde{\gamma}^1 ).
\ee
Here and in the following we denote all 2D quantities by a tilde.
For the 2D gamma matrices we use $\tilde{\gamma}^0 = \sigma_1$ and
$\tilde{\gamma}^1 = i \sigma_2$.  The 2D spinors are labelled by
$({s}_0)$.  Explicitly, $(\frac 12) = (1 \; 0)^T$ which corresponds to
a right moving fermion, and $(-\frac 12) = (0\; 1)^T$ corresponding to
a left moving fermion.

We dimensionally reduce $(s_0,s_1) \to (s_0)_{s_1}$, so that
right/left moving in 4D corresponds to right/left moving in 2D.  The
$s_1$ labels the charge under the global $U(1)$.\footnote{This can
formally be implemented by introducing a matrix $U =(1_2 \; \sigma_1)^T
$ such that $U^\dagger \Gamma^{0+} U = \tilde{\Gamma}^{0+}$.  Then
$\tilde{\psi} = U \psi$ is the same operation as $(s_0,s_1) \to
(s_0)$.  Helicity eigenstates, i.e., eigenstates of $\gamma^0 \gamma^3
\gamma^5 \psi = \pm \psi$, have $s_1 = \pm 1/2$.}  Applying this to
the zero mode wave function we get
\bea
\psi^0_+ &\to& \tilde{\psi}^0_+ 
= \alpha_+ {0 \choose 1}_{-1/2} + \beta_+  {0 \choose 1}_{+1/2}
\nonumber \\
\psi^0_- &\to& \tilde{\psi}^0_- 
= \alpha_- {1 \choose 0}_{+1/2} + \beta_-  {1 \choose 0}_{-1/2}
\label{zm_2D}
\eea
Each zero mode has 2 degrees of freedom (particle/antiparticle), and
can be described by a single 2D complex Weyl fermion.

The quantised Weyl spinor is
\be
\tilde{\psi}_W = \int_0^\infty \frac{\dd k} {2\pi} 
\(
\tilde{\psi}^0_n \tilde{a}(nk) \e^{-ik(z-nt)} 
+ \tilde{\psi}^0_n \tilde{b}^\dagger(nk) \e^{ik(z-nt)} 
\)
\ee
with $\tilde{\psi}^{0}_n$ given in Eq.~(\ref{zm_2D}).  We can now
compute the 2D conserved charges.  It is important to distinguish
between 2D and 4D chirality.  In D dimensions, chirality refers to
left and right-handed particles, eigenstates of $P_{L,R} = {(1_D \pm
\gamma^5_D) / 2}$.  The zero mode solution is a superposition of both
4D chiralities.  In 2D $\tilde{\gamma}^5 = {\rm diag}(-1,1)$.  Hence,
2D chirality is equivalent to left and right moving particles, which
is unrelated to the chirality concept in 4D. In 2D the vector and
axial currents are related $\tilde{j}^{\mu 5} = \epsilon_{\mu \nu}
\tilde{j}^\nu$.  The two conserved charges are
\be
\begin{array}{llll}
& \tilde{Q}_{n=1} = N_{\tilde{L}} - \bar{N}_{\tilde{L}} & {\rm and} &
\tilde{Q}_{n=1}^{5} = -(N_{\tilde{L}} - \bar{N}_{\tilde{L}}) \\
 {\rm or} & \tilde{Q}_{n=-1} = N_{\tilde{R}} -\bar{N}_{\tilde{R}}& 
{\rm and} & \tilde{Q}_{n=-1}^{5} = N_{\tilde{R}} -\bar{N}_{\tilde{R}}
\label{Q2D}
\end{array}
\ee
with $\tilde{L},\tilde{R}$ denoting 2D chiralities, i.e., left/right
moving. In 4D left/right moving is related to fermion getting their
mass via coupling to $\phi$ or $\phi^*$.  2D particle number
$\tilde{Q}$ corresponds to 4D particle number ${Q}$.  The 4D axial
current corresponds to the 2D global $U(1)$ current, under which
states with different $s_1$ are charged differently.  Furthermore, the
2D energy and momentum are equivalent to the 4D expressions.

The 2D Majorana fermion is obtained by setting $\alpha = \beta$ in
Eq.~(\ref{zm_2D}). It has 1 real d.o.f., corresponding to a
Majorana-Weyl fermion. The 2D charge conjugation matrix is $C = {\bf
1}_2$.  In addition, under charge conjugation also $s_1
\leftrightarrow -s_1$.  There is no global U(1) symmetry, no vector or
axial charges but there is a non-zero energy and momentum.

\subsection{Bosonisation}

If is well know that a 2D fermionic theory with Dirac fermions is
equivalent to a 2D bosonic theory. Here, we remind the reader how
fermionic superconductivity can be discussed in terms of bosonic
superconductivity \cite{Witten}.

If there are both left and right moving fermionic zero modes the
theory is anomaly free if the sum of charges of the left moving
particles is equal to those of the right moving particles.  A theory
with two zero modes of 4D Dirac particles moving in opposite
directions is effectively equivalent to two 2D Weyl fermions with
opposite chirality with the same charge.  These combine to form a
single 2D Dirac fermion. Such a theory can be bosonised. If the
fermion is charged with charge e, it is equivalent to a massive scalar
boson with mass $e/ \sqrt{\pi}$.

If on the other hand there is only a left (or right) moving Dirac zero
mode, which is necessarily neutral, the effective theory consists of a
single 2D Weyl fermion which has 2 real degrees of freedom. It is
equivalent to a free massless chiral boson with an action
\be
{\mathcal L}_B = \frac12 \partial_\mu \phi \partial^\mu \phi.
\label{boson1}
\ee
subject to the constraint:
\be
(\partial_t \pm \partial_z) \phi = 0.
\label{boson2}
\ee

The bosonic action in Eq.~(\ref{boson1}) is invariant
under $\phi \to \phi + n$, with $n$ integer (the integer comes from
the integer valued fermion current). Hence, $\phi$ is an angular
variable. A topological invariant can be constructed for a string loop
$Q = \int \dd l (\partial_z \phi)$~\cite{Witten}. The topological
invariant $Q$ corresponds to particle number $Q = N - \bar{N}$ of the
fermionic system. 

A Majorana zero mode has only one real d.o.f and can be described by a
2D Majorana-Weyl fermion.  Bosonisation is not possible, as a
Majorana-Weyl spinor has central charge $c=1/2$, whereas a chiral
boson has central charge $c=1$ --- the degrees of freedom do not
match. Vortons studies are commonly done in terms of the bosonic
description and it is not clear that this is adequate for a Majorana.

\subsection{Discussion}
\label{s:discussion}

The difference between the neutral Dirac current and the Majorana current is
the following.  

For the Dirac case particle number is non zero and flowing along the
string. It is a conserved quantum number. Current conservation is
guaranteed by an (accidental) $U(1)$ quantum number (such as lepton
number).  Two leptons cannot pair annihilate, and one lepton cannot
decay into a state without lepton number.  A left (right) moving zero
mode can be described by a left (right) moving Weyl fermion.  The
action can be bosonised, and is equivalent to a chiral boson.  One can
define a winding number which is conserved. This topological invariant
can be defined for the boson field which corresponds to particle
number in the fermionic description.

If there is only one fermionic zero mode the string is chiral.  This
can only happen for neutral fermions, otherwise there would be an
anomaly.  Although the current might be persistent, it does not
correspond to superconductivity, in the sense that there is no Meisner
effect. Moreover, in the fermionic language, there are no cooper
pairs.  Cooper pairs are bound states of fermions with opposite spin
and momenta, which therefore can only form if there are both left and
right moving currents~\cite{cooper}.

For the Majorana case, there is no chiral or vector charge. Thus
no-topological invariant can be defined. This is related to the fact
that a 2D theory with an odd number of Majorana-Weyl fermions cannot
be bosonised. Particle number is zero but in principle there can be a
non-vanishing number of particles: there is then an energy-momentum
flow.  Since the Majorana does not carry a conserved quantum number,
the number of Majorana excitations is not a conserved quantity in
general.  However, massless excitations cannot decay, neither pair
annihilate (as they do not scatter), and thus the current is stable
for kinematical reasons in the massless limit.

Note that the masslessness of the zero mode is only a good
approximation for an infinitely long straight string.  String
curvature and string oscillations will induce an effective mass.  This
will be discussed in more detail in the next section in the context of
vorton formation. The other issues we want to address is whether a non
negligible amount of chiral fermions can actually be trapped as
transverse zero modes on the string, and contribute to the string
energy-tension, and what is the actual difference between a flow of
Majorana-Weyl fermions and a current of neutral Weyl fermions.

\section{Vortons}
\label{s:vortons}

In this section we discuss fermionic superconductivity in cosmic
strings. The first ingredient we need is a fermionic zero mode. In the
previous section, we have discussed the zero mode in detail for both
Dirac and Majorana particles. Once we have a model which admits zero
mode solutions, we need to know whether these zero modes can be exited
or can be trapped so as to generate a current. Then finally, we must
check for the classical and quantum stabilities of these currents.

In the case of charged particles, a current can be generated as the
string moves through a magnetic field \cite{Witten}. This is not
possible for neutral particles. The only way that a large loop current
can be generated is if there is an initial tiny current. As the loop
collapses, the line integral of the current remains constant so that
the current grows as the inverse of the loop length. In the case of
charged particles, there are both left and right moving modes in order
to cancel anomaly; charged fermions can thus scatter and this process
strongly limits the current \cite{Barr}. In the case of a neutral
particle, there is only left or right moving modes, and the scattering
process is absent \cite{Barr}. Therefore large currents may only occur
if there is an asymmetry between left and right moving modes.  This
chiral asymmetry can only happen for neutral fermions which can either
be Dirac or Majorana.

Neutral currents cannot induce any (large scale) electric or magnetic
field.  Their cosmological interest lies solely in the conjecture that
fermionic currents can stabilise string loops against gravitational
collapse.  Such stable, superconducting string loops are called
vortons~\cite{vortons}.  Vortons contribute to the dark matter in the
universe.  The requirement that they do not overclose the universe can
put strong constraints on the underlying particle
physics~\cite{vorton_const}.

Vortons can be understood semi-classically
\cite{vortons,RLDavis}. Consider a fermionic current flowing along a
string loop.  Assume that the loop curvature does not significantly
alter the zero mode spectrum obtained in the limit of zero curvature.
To simplify all formulas, we further assume that for the Dirac current
$N \gg \bar{N}$.\footnote
{For $N \sim \bar{N}$ we find that the Fermi level is $\epsilon =
\max[N,\bar{N}]/R$ and fermionic energy $E_f = (N^2 +\bar{N}^2)/(2R)$,
results which only differ by order unity from those for $N \gg
\bar{N}$. In the literature vortons are often discussed in bosonised
language, in which the only quantum number is particle number
$Q=N-\bar{N}$, which can be substantially different from $N +
\bar{N}$.}
The zero mode energy levels are $\epsilon_n = n/R$, with $R$ the
radius of the string loop.  Each level can be occupied by at most
one fermion and one anti-fermion, in accordance with Pauli exclusion
principle.  In the ground state, the highest occupied level has Fermi
energy $\epsilon = {N}/R$.  The total fermionic energy is obtained by
summing over all occupied levels.  The total energy of string is the
sum of energy in bosonic and fermionic excitations $E = E_{\rm B} +
E_{\rm F}$:
\be
E = 2 \pi \mu R +  \frac{N^2}{2R}
\label{E_vorton}
\ee
where we have set $N(N+1) \to N^2$, valid in the limit $N \gg1$.
Further $\mu \sim \phi_0^2$ is the string energy per unit length.  As
the loops shrinks under gravitational action the bosonic energy
decreases, but the fermionic energy increases due to an increasing
spacing between energy levels.  The vorton radius, the radius for which
the energy is minimised, is
\be
R_0= \frac{N}{\sqrt{4\pi\mu}}
\label{R_0}
\ee
The above semi-classical description is valid for a loop radius larger
than the string width and than the Compton wavelength of the zero mode.
In particular, the last requirement, $R_0 > m_\psi^{-1}$, gives
\be
N > \lambda^{-1},
\label{vorton_valid}
\ee
with $\lambda$ the Yukawa coupling, see
Eqs.~(\ref{lag_D},~\ref{lag_M}).  We will discuss current formation
and stability for Dirac and Majorana currents, which may result in
vorton formation, in the rest of this section.

\subsection{Effective zero mode mass}

To understand vorton formation, to understand current formation and
stability, it is important to realise that the strings in the early
universe are not infinitely long straight strings.  Rather, the strings
vibrate and curve, there are cusps and intersections.  The zero-mode,
which is a strictly massless excitation (in the sense that it travels
at the speed of light) only for an infinitely long straight string,
obtains an effective mass under such condition.  Process which are
absent for massless zero modes, decay is an obvious example, become
possible.  It goes without saying therefore, that an effective mass
for the zero mode can alter the formation and stability of the
fermionic currents considerably.  Before considering these processes
in more detail, we will first discuss under which conditions the zero
mode gets effectively massive.

The zero-mode lives in a potential well of height $E = m_\psi$.  At the
bottom of the well the excitation is massless.  What keeps the zero
mode in a circular orbit for a string loop is that as the mode strays
off the centre of the vortex (bottom of the well) it becomes massive.
There is a restoring force (the centripetal force) $-\partial V/
\partial r \sim -\delta m/\delta r$ to keep it in orbit.  The
effective mass of the zero mode excitation can be estimated as
follows~\cite{Barr1}.  The acceleration of a particle going at the
speed of light around a circle with radius $R$ is $|\dot{k}| = k/R$.
Setting this equal to the centripetal force, and reminding that
$\delta r \sim m_\psi^{-1}$ with $m_\psi$ the vacuum fermion mass (since the
zero mode wave function falls off as $e^{-m_\psi r}$ at large $r$) we get
\be
m \sim \frac{\dot{k}}{m_\psi} \sim \frac{k}{R m_\psi}.
\label{curv_mass}
\ee
The higher the momentum of the mode the larger the effective mass.

Another way to see that string curvature induces an effective mass
(alters the dispersion relation to $E \neq |\vec{k}|$) is to consider
the solutions to the fermionic equations of motion.  Consider a string
loop with radius $R$ lying in the $xy$-plane, with its centre at the
origin.  Define also spherical coordinates ($r',\theta'$) which
parameterise the plane transverse to the string, with the origin at
the string core.  This plane is characterised by $\vec{r'} \times
\vec{\theta'} \parallel \vec{k}$, and $A_{\perp} = 0$.  The zero mode
solution is schematically of the form
\be
\psi = \e^{\pm i(E t + \vec{k} \cdot \vec{x})} 
\psi_0 (r',\theta').
\ee
Note that both $\vec{k}$ and $\vec{r'} \times {\vec{\theta'}}$ are
functions of time, since as the zero mode travels along the string
loop $\vec{k}/|\vec{k}|$ changes.  As a result the $\partial_t$
operator in the equation of motion not only brings down a factor $E$,
but also terms proportional to $\dot{k}/k$.  This gives a dispersion
relation of the form
\be
E \sim k + \frac{\dot{k}}{k} 
\label{disp}
\ee
Expanding $E = \sqrt{k^2 + m^2} \approx k + m^2 /k$ valid for $m \ll
k$, and setting $\dot{k} \sim m m_\psi$ equal to the centripetal force,
this gives the same mass correction as the classical argument leading
to Eq.~(\ref{curv_mass}).  

Another string configuration which leads to an effective mass for the
zero modes is a vibrating string.  Consider a string extended along
the $z$-direction, which moves with some velocity in the
$x$-direction.  The dispersion relation for the zero mode is $E^2 =
k_x^2 + k_z^2$.  By performing a Lorentz transformation we can go to a
frame in which the string is at rest, and $E = k_z$.  However, for a
vibrating string we cannot go to a frame in which the string is at
rest at all times, as $\dot{k} \neq 0$.  Just as for the case of a
string loop, the dispersion relation picks up a term $\dot{k}/k$, see
Eq.~(\ref{disp}).  For string loops the natural vibration frequency is
$\omega \sim 1/R$, whereas for string vibrations the natural scale is
$\omega \sim \sqrt{\mu} \sim \phi_0$.  The former case gives an
effective mass as in Eq.~(\ref{curv_mass}), whereas the latter gives a
mass which is larger by a factor $\phi_0 R$.

Finally, effective mass terms are generated if locally components of
the gauge field other than $A_\theta$ are excited, or if the Higgs
field profile deviates from its cylindric symmetric form.  The string
profile can be disturbed in string intersections or in cusps.  A
non-zero $A_z$ or $A_0$ component leads to a dispersion relation of
the form 
\be
E = k + q A,
\label{disp_A}
\ee
and the effective mass is $m^2 = q A k$.  A non-zero $A_r$ (or
non-cylindrical component of $\phi$) will appear as an effective mass
term in the $r,\theta$ dependent part of the equation of motion.  It
can be removed from the equation of motion by multiplying the zero
mode wave function by a factor $\e^{i q A z}$.  It thus likewise gives
a dispersion relation of the form Eq.~(\ref{disp_A}).  In cusps and
intersections $A \sim 1/r \sim m_f$ in Eq.~(\ref{disp_A}), and the
effective mass is large $m \sim k$.

The fermionic currents back react on the gauge field.  This creates an
effective mass for non-chiral strings but not for chiral
strings.~\cite{Ringeval1}

\subsection{Current formation}

We consider chiral strings only (as appropriate for Majorana currents)
and thus chargeless current carriers only.  There is no possibility of
current creation through spectral flow, as in the original Witten
model, as the fermionic zero modes are neutral.  Hence, any net
current must be the result of fermions getting trapped/condensing on
the string.  Currents may be created in thermal or non-thermal
settings.

\paragraph{Thermal trapping.} 
Let's first consider the case in which the string is emerged in a
thermal bath with a reheat temperature larger than the vacuum fermion
mass $T_R > m_\psi$.  Whatever the specifics of the trapping mechanism,
it is expected to be inefficient at high temperature $T \gg m_\psi$.  The
reason is that the typical energies involved in the trapping reactions
are much higher than the binding energy of the zero mode $E_b = -
m_\psi$, and trapping is suppressed.  Fermions that do get captured by
the string have typical momenta $k \gg m_\psi$ and can escape easily at
cusps and intersections. Furthermore, any zero mode present with $k
\lesssim m_\psi$ can be easily scattered off the string.  Hence, at high
temperature we expect the number of zero modes to be negligible small.
Current condensation has to take place at temperatures $T \lesssim
m_\psi$.

On the other hand, at temperatures $T \ll m_\psi$ capturing processes
are also suppressed.  For processes with one or more free fermions
$\psi$ in the in-state this is obvious, as their number density is
Boltzmann suppressed.  But also if the initial state contains only
light particles there is a suppressions factor.  The reason is that
the typical wavelength probed in the reaction is much larger than the
width of the zero mode, and the overlap between the zero mode wave
function and those of the initial particles is
small.  As a result reaction rates are smaller by
a factor $(E/m_\psi)^2$ compared to the rate expected on dimensional
grounds, with $E \sim T$ the typical interaction energy (see
appendix~\cite{bdm,dpd,ppdbm}).

We thus conclude that current formation, interactions in which zero
modes are captured/produced, has to take place at $m_\psi \sim T$.

Let us now consider the possible processes in which zero modes can be
produced.  The first interaction that comes to mind is a fermion
scattering off the string to produce a zero mode final state, that is,
the reaction $\psi \to \psi_{\rm zm}$ in the string background.  The
string is just a static background, it cannot exchange longitudinal
momentum or spin with the fermionic states.  Therefore energy and
longitudinal momentum cannot be conserved in the reaction if the mass
of the incoming fermion is larger than the effective mass $m$ of the
zero mode, see Eq.~(\ref{a:m_trap}).  And thus zero mode production in
fermion-string scattering does not occur.

But there are other reactions possible through which zero modes can be
produced.  If the only interactions present in the theory are those
given in Eqs.~(\ref{lag_D},~\ref{lag_M}), these are of the form $\psi
+ \psi \to \psi_{\rm zm} + \psi$.\footnote{An interaction such as
$\psi + \phi \to \psi_{\rm zm}$, possible at $T \sim m_\psi$ if
$m_\phi < m_\psi$ has infinitely small phase space as the spin of the
incoming fermion has to match that of the zero mode.}

In the limit $T \sim m_\psi$ the capturing and scattering rate are of
the same order of magnitude, and if the reaction is sufficiently fast,
a thermal equilibrium distribution of zero modes will be established.
We thus define the number of zero modes on a loop of length $L$ by
\be 
N \sim N + \bar{N} = p \frac{L}{\xi}.
\label{number}
\ee
with $\xi \sim T^{-1}$ the thermal correlation length. The efficiency
factor $p$ is of order unity if thermal equilibrium is reached, and 
$p\simeq 0$ if trapping reactions are frozen out
at $T \sim m_f$. The average particle number over many loops is zero
$\langle N - \bar{N} \rangle = 0$.  However, the variance, which gives
the order of magnitude of particle number on the individual loop is
non-zero: $\sqrt{ \langle (N - \bar{N})^2 \rangle } \sim \sqrt{N}$.

Thermal equilibrium can be established if the reaction rate is
sufficiently fast compared to the Hubble rate $\Gamma \sim n_\psi
\sigma > H$.  The thermal fermion density is $n_\psi \sim T^3 \sim
m_\psi^3$ at the temperature of interest.  The cross section for the
four-fermion interaction at energies $T \sim m_\psi$ is derived in
appendix~\ref{a:trapping}.  It is
\be
\sigma \sim g_X^4 A^6 \frac{m_\psi^2}{\max[m_X,\,m_\psi]^4}.
\label{sigma4}
\ee
As discussed in the appendix we expect the amplification factor $A$ to
be of order unity. If the reaction is mediated by the Higgs field $m_X
= m_\phi$ and $g_X = \lambda$ is the Yukawa, whereas if the reaction
is gauge mediated we have $m_X = m_A$ and $g_X = g$ the fermion gauge
coupling.  The reaction rate exceeds the Hubble rate for
\be
\frac{m_X}{g_X A^{3/2}} < (m_\psi^3 \mpl)^{1/4} = 
\left \{
\begin{matrix}
3 \times 10^{16} \GeV & {\rm for} \; m_\psi \sim 10^{16} \GeV \\
1 \times 10^{15} \GeV & {\rm for} \; m_\psi \sim 10^{14} \GeV \\
3 \times 10^{13} \GeV & {\rm for} \; m_\psi \sim 10^{12} \GeV 
\end{matrix}
\right.
\ee
If $m_X < m_\psi$, one should replace $m_X \to m_\psi$ in the above
equation.  If indeed $A \sim 1$ and $g_X \lesssim 1$ this becomes
impossible as $m_\psi \to 10^{16} \GeV$.

It is possible that the zero mode fermion $\psi$ couples to lighter
fermions $\chi$, for example, through an interaction $ \mcL_I = h
\hat{H} \hat{\bar{\psi}}_R \hat{\chi}_L + {\rm h.c.}$.  This is of the
form of the coupling of the right-handed neutrino to the left-handed
one. Such a coupling allows for trapping reactions $ \chi + \chi \to
\psi_{\rm zm} + \psi $ mediated by a light Higgs field.\footnote{The
reaction $f' + h \to f_{\rm zm}$ is phase space suppressed, as the
incoming light fermion needs to have the same spin as the zero mode.}
At energies $T \sim m_\psi$ (threshold energy) the cross section is
$\sigma \sim h^4 m_f^{-2}$.  This is of course of the same form as the
cross section Eq.~(\ref{sigma4}) in the limit $m_X < m_f$, and thus
the bound on $h$ derived below applies also to $g_X$ in this limit.
The trapping rate is $\Gamma \sim h^4 m_\psi$, which exceeds the
Hubble rate at temperatures $T \sim m_\psi$ for
\be
h  > \( \frac{m_\psi}{ \mpl} \)^{1/4} = 
\left \{
\begin{matrix}
0.3 & {\rm for} \; m_\psi \sim 10^{16} \GeV \\
0.1 & {\rm for} \; m_\psi \sim 10^{14} \GeV \\
0.03 & {\rm for} \; m_\psi \sim 10^{12} \GeV 
\end{matrix}
\right.
\ee
Large couplings are needed for effective current formation.

We conclude that thermal current condensation can only takes place at
$T \sim m_\psi$, and only if the trapping reactions are fast enough.
This last requirement is hard to satisfy in the limit $m_\psi <
m_A,m_\phi$ and in the absence of large Yukawa couplings to light
fermions.  If large Yukawa are present in the theory, current
formation is expected to happen for $m_\psi$ sufficiently small, with
the number of zero modes given by Eq.~(\ref{number}).

For Dirac spinors lepton number is substantially less than the the
total number of zero modes: $Q = N - \bar{N} \sim
\sqrt{N+\bar{N}}$.  Since massless particles do not annihilate, the
current description should be in terms of excitation rather than particle
number. In the literature, vorton formation is often discussed in
bosonised language in which the only quantum number is lepton number.
The bosonic language strictly only applies if all particle
anti-particle pairs have already annihilated, which would mean zero
current for the Majorana case.

\paragraph{Non-thermal capturing}
Fermionic current condensation can also happen non-thermally.  If
there is a large amount of energy released inside the string, this
will create bosonic excitations ($A$ and $\phi$), which decay into
fermions.  If decay is immediate, $\Gamma_{\rm dec} > H$ evaluated at
the time of the energy release, the bosons decay inside the string,
and the decay fermions may be captured as zero-modes.  This was first
discussed in~\cite{Barr2}, where they assume the energy-release is due
to an internal (only inside the string) phase transition.  This needs
a complicated Higgs sector, and we will not discuss it further.

One could also consider the case $T_R < m_\psi$.  Before decay the
Higgs field oscillates in its potential.  If $\dot{w} < w^2$ with $w =
k^2 + m^2_{\rm eff}(\phi)$ for all particles coupling to the Higgs
field ($\phi$, $A$, $\psi$), then the oscillation is adiabatic.  The
string profile function and the fermion zero modes change
adiabatically to adapt to the changing background fields.  However, in
the opposite limit, the evolution is non-adiabatically.  Parametric
resonance leads to explosive particle production.  The fermions thus
produced inside the string (from direct amplification, or from decay
of the bosonic fields) can be trapped.  It is not clear whether this
mechanism can be efficient.  One suppression factor is that only soft
quanta are exited during preheating with typical wave numbers $k_*$
much larger than the string width. Trapping is then 
suppressed by factors $(k_*/m_\psi)$.\\

To summarise, chiral currents which are necessarily neutral cannot be
created by spectral flow.  Current formation can still take place,
either thermally or non-thermally.  However, both processes are not
obvious, and require special circumstances.

\subsection{Current stability}

The effect of an (momentum dependent) effective mass opens up the
possibility for zero modes leaving the string, for pair annihilation,
and for zero mode decay.  We will discuss these processes in turn.

Classically, the zero mode cannot leave the string as long as $m <
m_\psi$. Conservation of angular momentum requires the massive mode
outside to take away all the momentum of the zero mode.  Consequently,
there is an energy barrier if $m < m_\psi$, but not in the opposite
limit.  As first noted in \cite{Barr1}, the generation of an effective
mass as in Eq.~(\ref{curv_mass}) puts a maximum on the current.  For a
vorton loop $m \sim k /(R_0 m_\psi)$ with $R_0$ given in
Eq.~(\ref{R_0}).  Using that the zero mode momentum is typically of
the order of the Fermi momentum $k_f \sim N/R_0$, the typical
mass is
\be
m \sim \frac{4\pi}{N} \frac{\phi_0^2}{m_\psi} \sim 
10^{14} \GeV \(\frac{10^5}{N}\) \( \frac{\phi_0}{10^{16} \GeV} \)^2 
\( \frac{10^{14}\GeV}{m_\psi}\).
\label{vorton_mass}
\ee
Requiring $m < m_\psi$ so that the current is classically stable gives
\be
\lambda^2 {N} > 4 \pi,
\label{lN}
\ee
where we have used that $m_\psi = \lambda \phi_0$. For perturbative
couplings this gives a stronger constraint than
Eq.~(\ref{vorton_valid}).  Although classically the zero mode cannot
leave the string if $m < m_\psi$, there is a non-zero tunnelling
probability for this process to happen~\cite{RLDavis}.  The extra
energy needed to overcome the energy barrier can be provided by the
string: enlarging the loop radius will increase the bosonic energy,
see Eq.~({\ref{E_vorton}).  The current is stable against tunnelling
processes (lifetime longer than age of the universe) for
~\cite{RLDavis}
\be
{N} \gtrsim \frac{100}{\alpha^{3/2}}
\ee
with $\alpha = g^2/(4\pi)$ the fine-structure constant of the broken
$U(1)_X$.

Another consequence of a non-zero mass is the possibility of zero mode
decay. As we have seen, thermal current formation is most effective if
the zero mode has a large, order unity, coupling to light particles.
Decay is kinematically allowed if the zero mode is heavier than the
decay products.  The decay rate for a coupling $h$, see
Eq~(\ref{a:decay}), then is
\be
\Gamma \sim \frac{h^2 m^3}{m_\psi^2} 
\( 1 + {\mathcal O}\(\frac{m}{m_\psi} \)^2 \).
\ee
For $m \ll m_\psi$ the Compton wavelength $m^{-1}$ is much smaller
than the width of the zero mode, and the wave function overlap of the
zero mode and the decay products is small. This gives a suppression
factor $(m/m_\psi)^2$.  For order one couplings and masses $m \gtrsim
10 \GeV$, decay occurs before big bang nucleosynthesis (BBN).  It
follows that current stability can only be assured if the decay is
kinematically forbidden and $m < m_{\rm dec}$.

A lower bound on the decay width comes from gravitational decay
\be
\Gamma \sim \frac{m^3}{\mpl^2} \(\frac{m}{m_\psi}\)^2
\ee
where we have again included a suppression factor $(m/m_\psi)^2$.
Decay happens before BBN, and thereby evading all bounds from vorton
overdensities, if
\be
m \gtrsim 10^8 \GeV \(\frac{m_\psi}{10^{14 \GeV}}\)^{2/5}.
\ee
This is generically satisfied for high scale phase transitions unless
${N}$ is extremely large.  Decay of Majorana particles cannot be
suppressed by invoking symmetries; decay of Dirac particles can be
suppressed by lepton number conservation, although it is believed that
gravity violates all global symmetries.

A third effect of a non-zero mass is that different modes travel at
different velocities and can scatter off each other.  Particles and
anti-particle states (Dirac case) or two particles can pair annihilate
(Majorana case) into decay products whose masses are lighter than the
Fermi energy. The typical centre of mass energy is of the order of the
zero mode mass.  Thus annihilation is only possible if the mass of one
of the zero modes is larger than the masses of the decay products. If
Eq.~(\ref{lN}) is violated, not only can zero modes leave the string,
also the reaction $\psi_{\rm zm} +\psi_{\rm zm} \to \psi + \psi$
becomes possible.  Scattering into lighter particles, $\psi_{\rm zm}
+\psi_{\rm zm} \to X$ is kinematically accessible for $m_X < m_\psi$.
The conditions under which scattering is effective are similar to
those under which zero mode decay occurs.  Note that pair annihilation
into light particles cannot be forbidden by lepton number.  If it does
happen, the Majorana current is destroyed, whereas a Dirac current can
subsist if there is a net lepton number.

To summarise, the zero-modes obtain an effective mass from string loop
curvature, string vibrations, and string perturbations making escape,
pair annihilation and decay of the zero modes possible.  Escape can be
prohibited for large enough ${N}$.  Annihilation and decay processes
might be suppressed or forbidden for Dirac fermions by conservation of
lepton number (though gravity is believed to violate global quantum
numbers), but not for Majorana fermions.  Decay and annihilation can
occur at relatively low Fermi levels if the zero modes couple to light
fermions --- which is precisely the favoured condition for thermal
current formation.  But even in the absence of light fermions,
gravitational decay happens before nucleosynthesis unless ${N}$ is
very large or the string forming phase transition happens at low
scale.  Vorton formation does not look probable, especially for
Majorana fermions.

\section{Conclusions}

In this paper we have derived the zero mode solution for a Majorana 
neutrino and compared it with its Dirac counterpart.  The Majorana zero 
mode has one degree of freedom, one less than the Dirac zero mode.   Both 
the vector and axial currents are zero.  It is therefore impossible to 
bosonise the 2D effective action for a Majorana zero mode.  

No conserved quantum number can be defined to assure the stability of a 
Majorana zero mode.  This is in contrast with Dirac fermions, for which 
always the analogue of lepton number can be introduced.  However, 
massless zero modes are stable for kinematical reasons.  And in the 
massless limits there is thus no difference between the stability of 
Majorana and Dirac zero modes.

The zero mode is only strictly massless for infinitely long straight 
strings.   The cosmic strings in our universe are expected to be curved, 
to vibrate, to fold and to intersect.   Under such conditions an 
effective mass for the zero mode is generated.  Needless to say that an 
effective mass alters the stability properties of a zero mode.   We have 
studied these issues in the context of vortons, string loops stabilised 
by a fermionic current.

Neutral fermionic currents can only be formed if zero modes are
somehow trapped on the string.  A massive fermion outside cannot
simply be trapped through scattering with the string, as this would
violate energy-momentum conservation. Interactions in which more than
two initial/final state particles participate are needed.  Such
interactions can only be efficient if the zero mode couples to light
particles with large, order one, couplings.

The fermionic current on a string loop increases as the loop contracts.  
The effective zero mode mass, which is proportional to the loop 
curvature, increases in this process, opening up the possibility of zero 
mode decay.  The decay width is large if the zero mode has a large 
coupling to light particles, precisely the conditions under which current 
formation is effective.  We expect that vortons will not form unless the 
string scale is low.   This is especially true for Majorana neutrinos 
which can pair annihilate, and whose decay cannot be protected by a 
conserved quantum number.

\section*{Acknowledgements}

We wish to thank Eugene Akhemedov, Lubo Musongela, Ilja Dorsner, Valery
Rubakov, Goran Senjanovic and Alexei Smirnov for usefull discussions.
RJ is supported by The Dutch Organisation for Scientific Research
[NWO].

\newpage

\appendix 
\section{Notation}

We use the Minkowski metric $\eta_{\mu \nu} = {\rm diag}(1,-1,-1,-1)$.
For the gamma matrices we take chiral basis:
\be
\gamma^\mu = 
\(\begin{matrix}
0 & \sigma^\mu \\
\bar{\sigma}^\mu & 0 
\end{matrix}\), 
\qquad 
\gamma^5 = 
\(\begin{matrix}
-1 & 0 \\
0 & 1 
\end{matrix}\)
\ee 
with $\sigma^\mu = (1, \sigma^i)$ and $\bar{\sigma}^\mu = (1,
-\sigma^i)$, and $\sigma^i$ the Pauli matrices: 
\be \sigma^1 =
\(\begin{matrix} 0 & 1 \\ 1 & 0
\end{matrix}
\), \qquad
\sigma^2 = \(\begin{matrix}
0 & -i \\
i & 0
\end{matrix}
\), \qquad
\sigma^3 =
\(\begin{matrix}
1 & 0 \\
0 & -1
\end{matrix}.
\) \ee  
In this basis, the projection operators $P_{\rm L,R}$ are defined by 
\be
\psi = \psi_{L} + \psi_{R} 
=  P_{\rm L} \psi + P_{\rm R} \Psi 
=  {{1 - \gamma^5}\over 2} \psi +  {{1 + \gamma^5}\over 2} \psi .
\ee 
The charge conjugate of a spinor is defined as
\be
\psi^c = C \psi^\dagger = \tilde{C} \bar{\psi}^T \; \; \; {\rm and} \;
\; \; \overline{\psi^c} = \psi^T \tilde{C}, 
\ee
with 
\bea 
C &=& i \gamma^2 ,\nonumber \\
\tilde{C} &=& C (\gamma^{0})^T = i \gamma^2 \gamma^0.
\label{CC} 
\eea
The latter matrix has the properties $\tilde{C} = -\tilde{C}^{-1} =
-\tilde{C}^\dagger =-\tilde{C}^T$.  The charge conjugate field
$\psi^c$ acts similarly as the hermitian conjugate field $\bar{\psi} =
\psi^\dagger \gamma^0$: it creates particles and annihilates
anti-particles.  The chiral projections of $\psi^c$ are given by
\be \psi^c_{L,R} = \tilde{C} \bar{\psi}^T_{R,L} 
\qquad {\rm and} \qquad
\bar{\psi}^c_{L,R} = {\psi}^T_{R,L} \tilde{C}
\ee %
Note that $\psi_L^c \equiv (\psi^c)_L = P_L \psi^c = P_L \tilde{C}
\bar{\psi}^T = \tilde{C} P_L \bar{\psi}^T = \tilde{C} (\bar{\psi}
P_L)^T = \tilde{C} \bar{\psi}_R^T = (\psi_R)^c$ is a left-handed
spinor, not to be confused with $(\psi_L)^c = (P_L \psi)^c =
\tilde{C} \bar{\psi}_L^T = (\psi^c)_R \equiv \psi^c_R$ which is a
right-handed spinor.

\section{Spinor conventions}

In this appendix we list our spinor conventions.  The normalization is
slightly different from that used in the paper.

\subsection{Free Dirac spinor}

We first give the results for a free massive Dirac spinor.  The
positive and negative frequency solutions of the Dirac equation are
\be
u(p,s) = 
{ \sqrt{p \cdot \sigma} \xi_s  
\choose
\sqrt{p \cdot \bar{\sigma}} \xi_s},
\qquad
v(p,s) = 
{ \sqrt{p \cdot \sigma} \xi_s  
\choose
-\sqrt{d \cdot \bar{\sigma}} \xi_s},
\ee
with $r =\pm$ corresponding the two independent
helicity/spin states.  In our conventions $\xi_+ = {1 \choose 0}$ and
$\xi_- = {0 \choose 1}$.  The spinors are normalized to
\bea
 u^\dagger(p,s) u(p,r) = & v^\dagger(p,s) v(p,r)& =
2 E_p \delta_{rs}, \\
 \bar{u}(p,s) u(p,r) = &-\bar{v}(p,s) v(p,r)& = 2 m \delta_{rs}, \\
 u^\dagger(p,s) v(-p,r) = &v^\dagger(-p,s) u(p,r)& = 0, \\
 \bar{u}(p,s) v(p,r) = &\bar{v}(p,r) u(p,r)& = 0.
\eea
The quantized field is
\be
\hat{\psi}(x) = \int \frac{\dd^3 p} {(2\pi)^3} \frac{1}{\sqrt{2 E_p}}
\sum_s \( 
\hat{a}_s(p) u(p,s) \e^{-i p \cdot x}
+ \hat{b}^\dagger_s(p) v(p,s) \e^{i {p \cdot x}} 
\).
\ee
Imposing an equal-time anti-commutation relation for $\hat{\psi}$
gives the anti-commutation relations for the creation and annihilation
operators:
\be
\{\hat{\psi}, \hat{\psi}^\dagger\} = \delta^3(\vec{x} - \vec{y}) 1_4
\quad \Longrightarrow \quad
\{ \hat{a}_s(p), \hat{a}^\dagger_r(k) \}
= \{ \hat{b}_s(p), \hat{b}^\dagger_r (k) \} 
= (2\pi)^3 \delta^3(\vec{p} - \vec{k}) \delta_{rs} 
\ee
with $1_4$ the unity matrix.  We define the one-particle state as
\be
|ps \rangle = 
\sqrt{2 E_p} \hat{a}^\dagger_s(p) |0\rangle.
\label{a:1p}
\ee
so that
\be
\langle p s| k r \rangle = 
2 E_p (2\pi)^3 \delta^3(\vec{p} - \vec{k}) \delta_{rs}
\label{a:1p_norm}
\ee
where we have used that the vacuum is normalized to unity $\langle 0|0
\rangle = 1$.  
%
%
Our renormalization choice Eq.~(\ref{a:1p}) corresponds to a
wave function normalization of $2E_p$ particles per unit volume. The
wave function is
\be
\langle 0 | \hat{\psi} | p s \rangle =
u(p,s) \e^{-i p \cdot x}
\ee

\subsection{Zero mode spinors}

The classical zero mode solution to the wave equation is (we consider
only $n=1$, and coupling to $\phi^*$)
\be
\psi^0(p) = 
{ \alpha \sqrt{p \cdot \sigma} \xi_- \choose
\beta \sqrt{p \cdot \bar{\sigma}} \xi^+}
\ee
This is the solution for both positive and negative frequencies.
Further we require 
\be 
\int \dd x \dd y \; (\alpha^2 + \beta^2) =1 
\label{a:norm_ab}
\ee
so that the spinors are normalized
\bea
\int \dd x \dd y \; {\psi^0}^\dagger(p) \psi^0(p) = 2 E_p \nonumber \\
\bar{\psi^0}(p) \psi^0(p) = {\psi^0}^\dagger(p) \psi^0(-p) =0
\eea
We approximate the profile functions by
\be
\alpha = - \beta = \frac{M}{\sqrt{\pi}} \e^{-M r},
\ee
satisfying the normalization condition Eq.~(\ref{a:norm_ab}).  The
zero modes for a vortex/anti-vortex with $|n| =1$ are independent of
the angular variable $\theta$.  The width of the wave function is set
by the fermion mass outside the string $M = m_\psi$\footnote{We
introduce $M$ so that in the interaction rate the origin of the
different factors can be traced easily.}, as the solution to the Dirac
equation at large $r$ is $\propto \e^{-m_\psi r}$.

The quantized zero mode field is
\be
\hat{\psi}_{\rm zm} = 
\int \frac{\dd p}{2 \pi} \frac{1}{\sqrt{2E_p}} 
\( \hat{a}_p \psi^0_p \e^{-ip(t-z)} 
+ \hat{b}^\dagger_p \psi^0_p \e^{ip(t-z)} \)
\ee
with $p \equiv |\vec{p}_L| = E_p$ with $\vec{p}_L$ the momentum
longitudinal to the string.  The field anti-commutator reads 
\be
\int \dd x \dd y \; \{ \hat{\psi},\hat{\psi}^\dagger \} 
\quad \Longrightarrow \quad
\{\hat{a}(p),\hat{a}^\dagger(k)\} 
= \{\hat{b}(p), \hat{b}^\dagger(k)\} 
= (2\pi) \delta^1_L(\vec{p} -\vec{k})
\ee
with $\bar{1}_4 = {\rm diag}(0,1,1,0)$.  The delta function involves
only the momentum aligned with the string. The matrix $\bar{1}_4$
instead of the unity matrix appears because the zero mode has only one
spin state, there is no summation over two independent spin states as
for the free Dirac field.  The one-particle state following from it
can be defined analogous to the free Dirac field
Eqs.~(\ref{a:1p},~\ref{a:1p_norm}).  The one-particle state is
\be
|p,s \rangle = 
\sqrt{2 E_p} \hat{a}^\dagger_s(p) |0\rangle.
\label{a:1p_zm}
\ee
so that
\be
\langle p s| q r \rangle = 
2 E_p (2\pi) \delta^1(p - k) \delta_{rs}
\label{a:1p_norm_zm}
\ee
The renormalization choice Eq.~(\ref{a:1p_zm}) corresponds to a
wave function normalization of $2E_p$ particles per unit length; it
reads
\be
\langle 0 | \hat{\psi} |p s \rangle =
\psi^0_{p s} \e^{-i p \cdot x} 
= u^0 \frac{M}{\sqrt{\pi}} \e^{-M r-i p \cdot x},
\ee
where we have defined $u^0$ through
\be
\psi^0 \equiv u^0 \alpha =  u^0 \frac{M}{\sqrt{\pi}} \e^{-M r}
\label{a:u_zm}
\ee
The product $u^0 \bar{u}^0$, which is needed in the calculation of the
squared amplitude $|\mcA|^2$, in the frame in which the string is
aligned with the $z$-axis and the zero mode has 4-momentum
$p^\mu=(p,0,0,p)$, is
\be
u^0 \bar{u}^0 = 2 p
\( \begin{matrix}
0 & 0 & 0 & 0 \\
-1 & 0 & 0 & 1 \\
1 & 0 & 0 & -1 \\
0 & 0 & 0 & 0 \\
\end{matrix} \).
\label{a:uubar}
\ee
Similar expressions exist for the anti-particle zero mode.

\subsection{Interactions in the string background}

The interaction rate for some processes involving a zero mode will be
calculated explicitly in the next two appendices.  Here we will give
some general remarks.

Following \cite{bdm,dpd,ppdbm} interactions in the string background
are calculated using the approximation that the incoming and outgoing
non-bound states are asymptotically free.  That is, we use a free
plane wave expansion for these states.  The plane wave expansion can
then be matched to an angular mode expansion, which is more natural in
the string background (as is done in Eq.~(\ref{a:g} below). As
the zero mode is bound to the string, it is a good approximation to
keep only the the lowest angular modes.  Possible amplification
factors, due to an amplification of the fermion wave functions of the
non-bound states near the string core, will be added by hand, see
Eq.~(\ref{a:A}) and further.

The main difference between interaction rates involving only
asymptotically free states, and interaction rates involving zero
modes are the following.

\begin{enumerate}
\item The zero mode has only one spin state. This gives factors of
order one difference, but does not lead to large quantitative or
qualitative differences.  
\item Conservation of transverse momentum is violated in the string
background.  For typical interaction energies $E \lesssim m_\psi$ this
has no big consequences.
\item The interaction rate is suppressed by a factor $(E /M)^2$ if the
typical interaction energy $E \lesssim M$ with $M \sim m_\psi$ the
width of the zero mode.  The reason is that the overlap between the
zero mode wave function, and the wave functions of the other states is
small~\cite{dpd}.
\item Non bound states can have wave functions which are enhanced near
the string core compared to the plane wave.  This factor is model
dependent as it depends on the core structure, the fractional flux and
the fermion charges.  This can lead to large amplifications of the
reaction rate~\cite{bdm,ppdbm}.
\end{enumerate}

\section{Zero mode decay}

To set the ground and to introduce the notation we start with a
discussion of the decay rate of a massless, asymptotically free
fermion.

\subsection{Decay of massless particle}

Consider the decay $\psi_R(k,r) \to H(p') \chi_L(k',r')$, mediated by a
term in the interaction Lagrangian
\be
\mcL_I = h \hat{H} \hat{\bar{\psi}} P_L \hat{\chi} + {\rm h.c.}
=  h \hat{H} \hat{\bar{\psi}}_R \hat{\chi}_L + {\rm h.c.}
\label{a:L_decay}
\ee
Decay is kinematically allowed only if the decay products are
massless, and if their momenta are collinear (and aligned with the
momentum of the decaying particle $\psi_R$). The phase space is
infinitesimal small. Since all products $p \cdot k$ vanish, with $p,k$
the 4-momenta of the particles involved in the decay process, 
all Mandelstam variables are zero.  The amplitude is
\bea
\mcA &=&
\int \dd^4 x \; \langle k'r';p' | \mcL_I  | kr \rangle
\nonumber \\
&=& (-ih) \int \dd^4 x \; \bar{u} (k',r') P_L u(k,r) \e^{-i(k-k'-p')}
\nonumber \\
&=& (2\pi)^4 \delta^4(k-k'-p') (-ih) {u}(k',r') P_L u(k,r).
\eea
Here $k$ is the momentum of the initial state fermion, $k'$ that of
the final state fermion, and $p'$ that of the final state boson.  We
use box normalization.  The interaction rate per unit volume can than
be defined through
\be
\dot{P}= |\mcA|^2/VT \equiv (2\pi)^4 \delta^4(k-k'-p') |\mcM|^2
\label{a:M}
\ee
with $ \mcA = (2\pi)^4 \delta^4(k-k'-p') i\mcM$.  The decay rate then
is
\be
\Gamma = \frac{1}{2E_k} \int \dd \omega_{k'}\dd \omega_{p'}
(2\pi)^4 \delta^4(k-k'-p') |\mcM|^2
\ee
with $\dd \omega_p$ the invariant 3-volume
\be
\dd \omega_p = \frac{\dd^3 p}{(2\pi)^3 2E_{p}}
\label{a:dwp}
\ee
Averaging over initial spin states and summing over final states, the
matrix element becomes
\bea
|{\mathcal M}|^2 &=& \frac{h^2}{2} \sum_{r,r'} 
{\rm Tr} \[ \bar{u}_{k' r'} P_L u_{k r} \bar{u}_{k r} P_R u_{k' r'} \]
\nonumber \\
&=& h^2 (k \cdot k') 
\nonumber \\&=&0
\eea
where the last equality applies for massless collinear particles.
Hence, the decay rate for a massless fermion is zero.

\subsection{Decay of the zero mode}

Consider now decay of the massless zero mode through the same channel
Eq.~(\ref{a:L_decay}). We have seen that the decay width for a massless
free particle is proportional to $k \cdot k'$, which vanishes for
massless collinear particles. Processes involving the zero mode do not
conserve momentum transverse to the string.  This is a consequence of
the fact that Lorentz symmetry (translations in the plane transverse
to the string) is broken by the presence of the string.  Therefore, the
decay products of the zero modes can acquire transverse momentum and
do not need to be collinear. Can this result in a non-zero decay width?

Let's calculate.  The matrix element for $\psi_{\rm zm}(k,r) \to
\psi(k',r')$ is
\bea
\langle k' r' | \bar{\psi} P_L \psi_{\rm zm} | kr \rangle
&=& \bar{u}_{k'r'} P_L \psi^0 \e^{-i(k-k')_L x_L} \e^{-i k'_T x_T}
\nonumber \\
&=& 
\bar{u}_{k's'} P_L u^0_{k} \e^{-i(k-k')_L x_L} 
\frac{M}{\sqrt{\pi}} \e^{i k'_T x_T-M r}
\eea
where in the last line we have used Eq.~(\ref{a:u_zm}).  The subscript
$L$ and $T$ denote space-time longitudinal (including the
time-coordinate) and transverse to the string respectively.  For the
zero mode $k_T = 0$.  The decay amplitude for the process $\psi_{\rm
zm}(k) \to H(p')\psi(k',s')$ then is
\bea
{\mathcal A} 
&=& -i h \int \dd^4 x 
\langle k' r' | \hat{\bar{\psi}} P_L \hat{\psi}_{\rm zm} | k r \rangle
\langle p'|\hat{H}|0\rangle
\nonumber \\
&=& (2\pi)^2 \delta^2_L(k-k'-p') (-ih)\bar{u}_{k'r'} P_L u^0_{kr}  
\frac{g(q_T)}{M}
\eea
with $q_T = (k'+p')_T$.  In the last line we have introduced the
dimensionless function $g$ through
\bea
\frac{g(q_T)}{M} &=& \int \dd^2 x_T \;
\frac{M}{\sqrt{\pi}} \e^{i k'_T x_T-M r}
\nonumber \\
&\approx& \frac{M}{\sqrt{\pi}} 2\pi \int \dd r r J_0(q_T r) \e^{-Mr} 
=  \frac{2\sqrt{\pi}}{M} (1+q_T^2/M^2)^{-3/2}
\label{a:g}
\eea
where in the first step we have only kept the lowest angular mode.
There is no delta function imposing conservation of transverse
momentum.  Rather the integration over transverse coordinates lead to
a distribution which has maximum $g \sim 2\sqrt{\pi}$ for $q_T =0$ and
decreases (power law) for $q_T \gg M$. There is a non-zero probability
for the decay products to acquire transverse momentum.

Using box normalization we can define the interaction rate per
unit length as
\be
\dot{P} = |\mcA|^2/LT = (2\pi)^2 \delta_L^2(k-k'-p')|\mcM|^2.
\ee
The decay width then is
\be
\Gamma = \frac{h^2}{2E_k} \int \dd \omega_{k'} \dd \omega_{p'}
(2\pi)^2 \delta^2_L(k-k'-p') \( \frac{g(q_T)}{M}\)^2 
\sum_{r'}|\bar{u}_{k'r'} P_L u^0_{k}|^2
\label{a:dec}
\ee
The delta function in the decay width can only be satisfied for
massless decay products, with their longitudinal momentum in the same
direction as the zero mode momentum.  In particular, it implies
$E_{k'} =k'$ and $q_T = 0$.  Since the decay product are then
collinear just as in the decay of a non-bound fermion, we also expect
the zero mode decay width to vanish. Indeed, the spinor factor,
evaluated in the frame where the string is aligned with the $z$-axis
so that the zero mode momentum is $k^\mu=(k,0,0,k)$, is
\be
\sum_{r'} |\bar{u}_{k'r'} P_L u^0_{kr} |^2 =  
\Sigma_{s'} {\rm Tr} \[ \bar{u}' P_L u^0 \bar{u}^0 P_R u' \]
= 2k (E_{k'}- k'_z)
= 0,
\ee
where in the second step we have used Eq.~(\ref{a:uubar}), and the
equivalent for the massless free spinor $\Sigma_{s'} u' \bar{u}' =
\slashed{k}'$.  

Energy-momentum conservation in the longitudinal plane assures that a
massless zero mode does not decay.

\subsection{Decay of massive bound state}

If the zero mode gets an effective mass $m$ no longer $E_k = k$, and
thus no longer necessarily $E_k' = k'$, and the decay width can be
non-zero.  We will assume that the zero mode spinor is unaffected by
the effective mass, only the frequency part is altered.  The cross
section is most easily evaluated in the center of mass frame, where
the zero mode is at rest.  The spin sum give $\sum_{s'}
|\bar{u}_{k's'} P_L u^0_{ks} |^2 \sim m^2$. Introducing an extra delta
function the decay width Eq.~(\ref{a:dec}) then becomes
\bea
\Gamma &\sim& \frac{h^2 m}{M^2} \int \dd \omega_{k'} 
\dd \omega_{p'} \dd^2 q_T\;
(2\pi)^2 \delta^4_L(k+q_T-k'-p') g(q_T)^2 
\nonumber \\
&\sim& \frac{h^2 m}{M^2} \int_0^{m} \dd^2 q_T \; g(q_T)^2 
\nonumber \\
&\sim& \frac{h^2 m^3}{M^2} \( 1 + {\mathcal O}\(\frac{m}{M} \)^2 \).
\label{a:decay}
\eea
Here, as before, $q_T = (k'+p')_T$ is the transverse momentum provided
by the string.  The integration over $k'$ and $p'$ can be done in the
usual way, and impose energy-momentum conservation in the longitudinal
plane.  Energy conservation restricts $q_T \lesssim m$.  In the last
step we have taken the limit $m \ll M \sim m_\psi$, with $m_\psi$ the
fermion vacuum mass, to arrive at the final result.  For $m \ll M$
there is a wave function suppression.  The reason is that then the
Compton wavelength $m^{-1}$ is much smaller than the width of the zero
mode.

\section{Zero mode trapping}

We are interested in processes with in the initial state free
fermion(s), and in the final state a zero mode.  If such reactions are
fast compared to the Hubble rate at $T \sim m_\psi$, zero modes are
trapped, and chiral currents form.

\subsection{Zero mode from string scattering}

We first consider the process 
\be
\psi(k,r) \to \psi_{\rm zm} (k',r')
\label{ffzm}
\ee
in the background of the string.  The interaction can take place via
either the string Higgs or gauge field.  As we will see, the reaction
rate is zero as a consequence of energy-momentum conservation.

The amplitude for $\psi(k,r) \to \psi_{\rm zm} (k',r')$ through a
fermion-Higgs coupling of the form
\be
\mcL = \lambda \hat{\phi} \hat{\bar{\psi}} \hat{\psi}.
\ee
is
\be
\mcA = \int \dd^4 x \langle S';k'r'| \mcL_I | S;kr \rangle
=\int \dd^4 x 
\langle k'r' | \hat{\bar{\psi}}\hat{\psi} | kr \rangle
\langle S'| \hat{\Phi} | S \rangle
\ee
with $S$ denoting the string state.  If the back-reaction is small, if
the string is not much altered by the scattering, then $S' \approx S$.
Here $\hat{\Phi} = \hat{\phi} - \phi_0$, the field shifted by the VEV
outside the string.  We approximate 
\be
\langle S'| \hat{\Phi} | S \rangle  = 
\left\{ \begin{matrix}
 \phi_0, & \quad r < m_\phi^{-1}, \\
 0, & \quad r > m_\phi^{-1}.
\end{matrix} \right.
\ee
The amplitude then becomes
\be
\mcA \sim  (2\pi)^2 \delta^2_L(k-k') (-i \lambda)
\bar{u'}^0 u \delta_{r r'} \frac{M \phi_0}{m_\phi^2} 
\label{mca_string}
\ee
where we have used that 
\bea
\int_0^{m_\phi^{-1}} \dd^2 x_T \phi_0 \frac{M}{\sqrt{\pi}} 
\e^{-ik_T \cdot x_t - M r}
&\sim& M \phi_0 \int_0^{m_\phi^{-1}} r \dd r J_0(k_T r) \e^{- M r}
\nonumber \\
&\sim& \frac{M \phi_0}{m_\phi^2} \( 1 + {\mathcal O}
\(\frac{k_T}{m_\phi},\frac{M}{m_\phi} \) \)
\eea
The integration over longitudinal momentum gives a delta function
enforcing energy-momentum conservation in the longitudinal $(t,z)$
plane.  The string is just a stationary background, with no
longitudinal string momentum.  Note however, that the string can
provide transverse momentum so that $q_T \lesssim E$ can be non-zero,
with $E$ the typical interaction energy.  

Energy momentum conservation in the longitudinal plane, which reads
explicitly $E_k=E_k' = k'$ and $k_3 = k'_3=k'$, is only possible if
the incoming fermion is also massless.  The same conclusion follows
from the Dirac delta in Eq.~(\ref{mca_string}).  The spin of the
incoming fermion should be equal to the spin of the zero mode fermion.
Only for massless incoming fermions is this possible. The massless
spin states outside the string are chiral eigenstates and only in this
case is it possible to write the zero mode wave function $\psi^0$ as a
superposition of the fermion states outside the string.\footnote {The
zero mode spinor is of the form $u^0 = \sqrt{2k}(0 \; 1\; 1 \; 0)^T$
and the two massless free states are $u_+ = \sqrt{2k} (0 \;0 \; 1
\;0)^T$ and $u_- = \sqrt{2k} (0\; 1\; 0\; 0)^T$.  It follows $u^0 =
u_- + u_+$. } We conclude that the process Eq.~(\ref{ffzm}) to capture
zero modes is impossible if the fermions are massive outside the
string.

We will show here that the cross section is also zero for massless
incoming fermions. The cross section per unit length is
\be 
\frac{\dd \sigma}{\dd l} \sim \frac{\lambda^2}{E} 
\int \frac{\dd k'}{(2\pi)2k'}
\delta^2_L(k-k')\frac{1}{2}\delta_{rr'} |\bar{u'}^0 u|^2  
\( \frac{M \phi_0}{m_\phi^2}\)^2.
\ee
Evaluating the spin term in the frame where the zero mode momentum is
$k'^\mu = (k', 0, 0,k')$ gives
\bea
\delta^2_L(k-k')  \delta_{rr'}  |\bar{u}^0(k') u(k,s)|^2 
&=&\delta^2_L(k-k') 2k' (E_k - k_3)
\nonumber \\
&=& 0
\eea

String scattering through the interaction term $\mcL = iq
\hat{\bar{\psi}} \gamma^\theta \hat{A}_\theta \hat{\psi}$ gives
similar expression.  It can easily be checked that also in this case
the cross section is zero.

Can string scattering lead to zero mode capture if the zero mode is
effectively massive?  Energy momentum conservation in the longitudinal
plane gives $k_T^2+m_\psi^2 = m^2$, with $k_T$ the transverse momentum of
the incoming particle, $m_\psi$ the vacuum mass, and $m$ the effective
zero mode mass.  It follows that zero mode capture through
Eq.~(\ref{ffzm}) is impossible for
\be
m_\psi > m
\label{a:m_trap}
\ee
which is the case of interest.

\subsection{Four fermion scattering}
\label{a:trapping}

Consider 
\be 
\psi(k,r) + \bar{\psi}(p,s) \to \psi_{\rm zm}(k') + \bar{\psi}(p',s')
\ee
mediated by either a Higgs or gauge boson.  Since the zero mode wave
function For simplicity we consider only s-channel Higgs mediation
explicitly; other diagrams and interference terms give similar
contributions.  The amplitude for this process is
\bea
\mcA &=& (-i \lambda)^2 \int \dd^4 x_1 \dd^4 x_2
\bar{v}(p,s) u(k,r) \frac{1}{q^2 + m_\phi^2} \bar{u}^0(k') v(p',s') 
\e^{-i(k+p-q) \cdot x_1 -i (q -k' -p') \cdot x_2 - M r}
\nonumber \\
&\sim& (2\pi)^2 \delta^2_L(k+p-k'-p')  \frac{(-i \lambda)^2}{m_\phi^2} 
\bar{v}(p,s) u(k,r) \bar{u}^0(k') v(p',s') \frac{g((q-p')_T)}{M}
\eea
where we have assumed that the interaction takes place a distance $r <
m_f^{-1}$ away from the string core.  In the opposite limit $r \ll
m_f^{-1}$ the zero mode wave function approaches zero exponentially
fast, and the cross section is zero.  In the second line we have
specialized to interaction energies $q = k+p \ll m_\phi$.  The
function $g$ is defined in Eq.~(\ref{a:g}).  The cross section (for
interactions taking place within a distance $r < m_f^{-1}$ from the
string center, is
\be
\sigma \sim \frac{\lambda^4}{m_\phi^4} 
\int \dd \omega_{p'} \dd \omega_{k'} \dd^2 q_T \; \delta^4_L(q-k'-p') 
\frac{1}{4}\Sigma_{r,s,s'}
|\bar{v} u \bar{u}^{0'} v'|^2 \frac{g(q_T)^2}{M^2}
\ee
with $ \dd\omega_{p'}$ given in Eq.~(\ref{a:dwp}), and $\dd
\omega_{k'} = \dd k/(4\pi k)$.  The spin sum is evaluated in the frame
in which the zero-mode momentum is $k'^\mu = (k',0,0,k')$:
\be
\frac{1}{4}\Sigma_{r,s,s'}|\bar{v} u \bar{u}^{0'} v'|^2
= 4k'\((k_0-k_3)(m^2-p \cdot p')-(p_0'-p_3')(m^2-k \cdot p) \)
\ee
which is non-zero for massive fermions (so that $|k_0| \neq |k_3|$ and
$|p'_0| \neq |p'_3|$). For interaction energies $q \sim m_\psi$ the cross
section then is
\be
\sigma \sim \lambda^4 \frac{m_\psi^4}{m_\phi^4 M^2} 
\int \dd^2 q_T g^2
\stackrel{M \sim m_\psi}{\longrightarrow} 
\lambda^4 \frac{m_\psi^2}{m_\phi^4}
\label{a:sigma4}
\ee
For $m_\psi \ll M$ there is a $(m_\psi/M)^2$ suppression as the Compton
wavelength $m_\psi^{-1}$ is much smaller than the width of the zero mode.
In this calculation we have use that the free particles are of the
plane wave form.  However their wave function may be amplified on the
scale of the typical interaction length $m_\psi^{-1}$.  Defining
\be
A = \frac{|\tilde{u}(r=m_\psi^{-1})|}{|u(r=m_\psi^{-1})|}
\label{a:A}
\ee
with $\tilde{u}$ the wave function in the string background and $u$, as
before, the plane wave.  The cross section corrected by these
amplification factors becomes
\be
\sigma_{\rm amp} \sim A^6 \lambda^4 \frac{m_\psi^2}{m_\phi^4}.
\ee
The amplification factor depends on the core profile and on the
fractional flux.  We expect the amplification factor to be
\be
A \sim \( \frac{1}{kr} \)^p \lesssim \( \frac{m_\psi}{k} \)\sim 1
\ee
with the coefficient $p \leq 1$.\footnote{In \cite{ppdbm} string
scattering is discussed and $A$ is evaluated at $r \sim m_\phi^{-1}$,
resulting in a factor which can be as large as $(m_\phi/k)$.  The four
fermion interaction we discuss takes place at $r\sim m_\psi^{-1}$ and we
expect amplification to be minimal.}

\end{document}